\documentclass[prl,a4paper,twocolumn,showpacs,superscriptaddress,amssymb]{revtex4}
\usepackage{graphicx}
\usepackage{hyperref}

\begin{document}

\title{Universal Critical Velocity for the Onset of Turbulence of Oscillatory Superfluid Flow}

\author{R. H\"anninen}
\affiliation{Low Temperature Laboratory, Helsinki University of Technology, FIN-02015 TKK, Finland} 
\author{W. Schoepe}
\affiliation{Institut f\"ur Experimentelle und Angewandte Physik, Universit\"at Regensburg, D-93040 Regensburg, Germany} 
\date{\today}

\begin{abstract}  
The critical velocity $v_c$ for the onset of turbulent drag of small spheres oscillating in superfluid $^4$He is frequency dependent ($\omega/2\pi$ from 100 Hz to 700 Hz) and is described by $v_c$ = 2.6$\,\sqrt{\kappa \,\omega}$, where $\kappa$ is the circulation quantum. A qualitative analysis based on a recent theory of the onset of superfluid turbulence gives $v_c  \approx \sqrt{8\kappa \,\omega/\beta}$, where $\beta\sim1$ depends on the coefficients of mutual friction. This agrees well with the data and implies that $v_c$ is a universal critical velocity that is independent of geometry, size, and surface properties of the oscillating body. This is confirmed by comparing our data on spheres with $v_c$ obtained with other oscillating structures by other groups. Numerical simulations indicate somewhat larger critical velocity, above which a rapid increase in vortex length is observed.  
\end{abstract}

\pacs{67.25.dk, 67.25.bf, 47.27.Cn}

\maketitle

\section{Introduction}\label{intro}
A quantitative description of the critical velocity for the onset of turbulence caused by a macroscopic body moving in superfluid helium is still an unsolved problem. At low velocities the superfluid flowing around the body exhibits pure potential flow and the only drag force is due to scattering by thermally excited quasiparticles (rotons and phonons in $^4$He) whose density rapidly vanishes when the temperature is decreased to the millikelvin regime. If, however, the velocity is increased a sudden onset of a large and nonlinear drag force is observed at a critical velocity $v_c$. This is due to a breakdown of potential flow and the onset of turbulence due to vortex production by the moving body. These vortices all have the same quantized circulation $\kappa = h/m \approx 10^{-7}$ m$^2$/s ($h$ is Planck's constant and $m$ is the mass of a $^4$He atom) and a core of atomic size. These vortices form a tangle that is described as ``superfluid turbulence" or ``quantum turbulence".
  There is a large body of experimental evidence that $v_c$ depends on the size of the moving body \cite{Donnelly}. While this is true for steady flow, the question arises whether the physics will be different for oscillatory flow. After all, for practical reasons, often an oscillating body like a sphere, a vibrating wire, a small tuning fork, or even a vibrating grid are employed in the experiments \cite{Vinen}. 

In our present work, we first show experimentally that in the case of oscillatory flow generated by an oscillating body there is a universal critical velocity, that is given only by the oscillation frequency $\omega$, the circulation quantum $\kappa$, and a dimensionless number of order one, {\it i.e.}, $v_c\sim\sqrt{\kappa \cdot \omega}$. This result can be made plausible simply on dimensional grounds \cite{Yano0,arxiv}. Finally, an analysis based on a rigorous theory for the onset of superfluid turbulence in steady counterflow, when extended qualitatively to the case of oscillating flow in our experiments, gives good agreement with the data, including their very weak temperature dependence from millikelvin temperatures up to 1.9 K. 

\section{Experiments}\label{experiments}

\begin{figure}[ht]
\centerline{\includegraphics[width=0.9\columnwidth,clip=true]{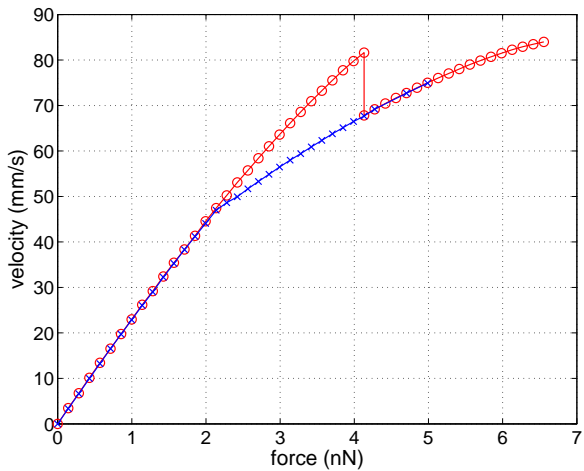}}
\caption{\label{f.veloVSforce} (color online) Velocity amplitude as a function of the driving force for a 100 $\mu$m sphere oscillating at 236 Hz in superfluid $^4$He at 1.90 K. Red circles: data taken with increasing drive; blue x :  decreasing drive. Because of the strong hysteresis $v_c$ = 46 mm/s can only be determined with the down sweep.}
\end{figure}

The experiments with spheres were performed at Regensburg University several years ago. Most of the results have been published in earlier work \cite{intermitt1,physica,PRL1}, but a systematic analysis of the frequency dependence of the critical velocities was only performed recently. The data were taken with 2 different spheres, having a radius of 100 $\mu$m and of 124 $\mu$m. The smaller one was used in a $^3$He cryostat, the other one in a dilution cryostat. The magnetic spheres were placed inside a horizontal parallel plate capacitor consisting of niobium electrodes to which a voltage was applied while the temperature was dropped below the superconducting transition temperature of niobium, ca.10 K. At helium temperatures the charged spheres were levitating between the capacitor plates. Frozen flux in the electrodes was necessary for a stable vertical oscillatory motion when a small ac voltage was applied at resonance. The frequency of the oscillators could not be controlled quantitatively because it depended on the frozen flux of the particular levitation status. It was observed, however, that a faster cooling rate from 10 K to 4.2 K in general lead to higher resonance frequencies. The critical velocities were measured by reducing the driving ac voltage in the turbulent flow regime (identified by the large nonlinear turbulent drag) until turbulence vanished. This procedure was necessary because if, instead, the drive was increased from zero, it was found that often the critical velocity could be substantially exceeded before potential flow eventually would break down \cite{PRL1}, see Fig.~\ref{f.veloVSforce}. The origin of this hysteresis will be discussed in the following Chapter.

\begin{figure}[ht]
\centerline{\includegraphics[width=0.9\columnwidth,clip=true]{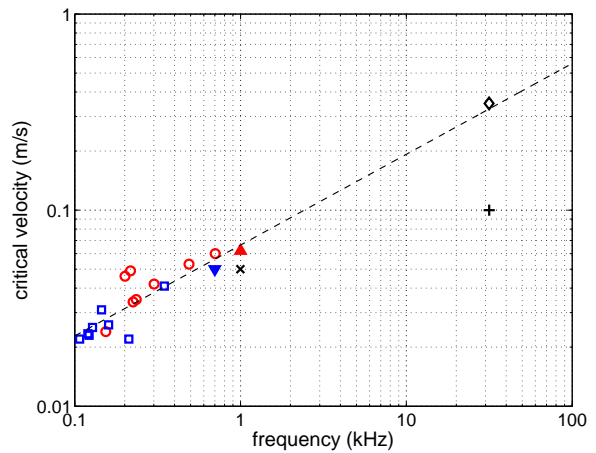}}
\caption{\label{f.vcsphere} (color online) Critical velocity for the onset of turbulence in superfluid $^4$He as a function of the frequency. Blue squares: larger sphere; red circles: smaller sphere, temperature is below 1 K; red triangle: vibrating wire at 7 mK \cite{Pick}; blue inverted triangle: vibrating wire at mK temperatures \cite{Yano}; x : vibrating grid at 8 mK \cite{grid}; diamond: tuning fork at 1.5 K \cite{Hosio}; + : tuning fork at 1.3 K \cite{Ladik}. The dashed line is a fit to all data (except for the +): $v_c=2.0\,(\kappa 2\pi\!f)^{0.46}$.} 
\end{figure}

At temperatures below 0.5 K the hysteretic behavior is replaced by an instability of turbulence leading to an intermittent switching between potential flow and turbulent phases whose lifetimes gradually vanish when $v_c$ is approached by reducing the drive \cite{intermitt1}. This instability has been attributed to statistical fluctuations of the length of vortex lines per unit volume $L$ \cite{intermitt2}. The resulting $v_c$ of the various runs for both spheres are displayed in the Fig.~\ref{f.vcsphere} together with critical velocities obtained with other oscillating structures by other groups.

We first discuss our own results for the spheres. The data in Fig.~\ref{f.vcsphere} were all taken below 1 K where the critical velocities were found to be independent of temperature, see below.  
The errors of the individual values for the spheres vary from ca. 1\% to about 5\%, depending on how closely $v_c$ was approached in the experiments. The exceptionally large scatter of the three data points near 200 Hz is by far larger than any error bar, the reason of which is not clear: only a small systematic error could have occurred in the calibration of the velocity scale (caused by a small error of few percent in the determination of the electric charge of the oscillating sphere). The mostly higher resonance frequencies of the smaller sphere are due to its smaller mass and probably also due the usually faster cooling rate from 10 K used with the $^3$He cryostat. 

The other data we could find in the literature are the following. 1. A result from an experiment with a vibrating wire of the Lancaster group \cite{Pick}. In that experiment it was found that at the onset of turbulence the resonance frequency would increase and from their figure we obtain $v_c\approx$ 62 mm/s. 2. A citation by the Osaka group \cite{Yano} of their result with a vibrating wire. (This group has recently obtained more data \cite{Yano1}.) 3. One result on vibrating grids $\approx$ 50 mm/s \cite{grid}. We have no information on the accuracy of those data. It should be mentioned that because the velocity amplitude of these structures varies from zero at the fixed ends to a maximum in the middle, the measured velocity signal is some average from which the maximum velocity has to be calculated by some model. The resulting data are fairly consistent with ours. 3. One result with a tuning fork of the Helsinki group \cite{Hosio} and one from Prague \cite{Ladik}. The calibration of the velocity scale at the tip of a tuning fork has been presented in Ref.\cite{Blau}. We note that the value from Prague is lower than the result from Helsinki by a factor of $\sim$3.5. The origin of this discrepancy is not clear, we can only speculate that signal losses in the cables might have occurred and remained undetected. Therefore, at present this data point remains uncertain to us and is not included in the fit. Fitting a power law to all the other data gives $v_c = 2.0\,(\kappa\,\omega)^{0.46}$ (in SI units). It is obvious that this fit gives convincing evidence of the scaling $v_c\sim\sqrt{\kappa\,\omega}$ that was inferred before from purely dimensional considerations. If we fix the power to 0.5 we get as the final result $v_c\approx$ 2.6\,$\sqrt{\kappa \,\omega}$. It should be mentioned that a very recent work by the Osaka group on wires vibrating at various frequencies also shows the same scaling of $v_c$, including the frequency range from 1 kHz to 10 kHz where we have no data \cite{Yano1}. 

\section{Qualitative model}\label{model}
The dimensional considerations leading to the frequency dependence of $v_c$ are supported by the following qualitative but very general argument. A ``superfluid Reynolds number" $Re_s = vl/\kappa$ has been defined, where $l$ is some characteristic length scale \cite{sRe,Kolya}. For $Re_s>$1 turbulence is expected to be stable when mutual friction is negligible. It is reasonable to assume that for oscillation amplitudes $a = v/\omega$ that are small compared to the size of the sphere, $a$ ($\le$ 30 $\mu$m) is the characteristic length scale. Hence, we may expect to observe turbulence for $v \ge v_c \sim \sqrt{\kappa\,\omega}$ .

Another qualitative theoretical analysis is based on comparing the time scale of the oscillatory flow $\omega^{-1}$ with the time scale governing the dynamics of the vortex tangle. In a recent work on the onset of superfluid turbulence in {\it steady} counterflow by Kopnin \cite{Kolya} a rate equation for the length of vortex lines per unit volume $L$ (m$^{-2}$) has been derived:
\begin{equation}
\dot{L} = \beta[v_s\,L^{3/2} - \kappa\,L^2],
\label{e.Ldot}
\end{equation}
where $\beta = A(1-\alpha^{\prime}) - B\alpha$ is determined by the coefficients of mutual friction $\alpha^{\prime}$ and $\alpha$ and the constants $A, B \sim$ 1 (more precise values for $A$ and $B$ would require numerical simulations with some particular geometry), and $v_s$ is the velocity of the superfluid (with respect to the quasiparticle gas).  In superfluid $^4$He $\beta \sim$ 1 except very near the Lambda transition. Equation~(\ref{e.Ldot}) is a generalization of the well known Vinen equation \cite{VinenEq} that determines the balance between the rate of growth of $L$ due to vortex multiplication and the losses due to mutual friction with the quasiparticles. Stable solutions of Eq.~(\ref{e.Ldot}) are $L$=0 and the saturated value $L_{sat} = (v_s/\kappa)^2$. An analytical solution for $\beta>$ 0 and constant $v_s$ can be found in \cite{Schwarz}. A simple approximation is derived by Kopnin for the limit of an initial growth from a small value $L_0 \ll L_{sat}$ at $t$=0: 
\begin{equation}
L/L_{sat} = [(L_{sat}/L_0)^{1/2} - (t/\tau)]^{-2},
\label{e.Lgrow}
\end{equation}
where the time constant $\tau = 2\kappa/\beta v_s^2$ sets the time scale for the initial growth. After a time $t\sim\tau\sqrt{L_{sat}/L_0}$ a very rapid growth (``vortex avalanche") sets in. Finally, the stable saturated value $L_{sat}$ is approached exponentially: $L_{sat}-L\sim \exp(-t/\tau)$.

In order to extend this scenario to the case of {\it oscillatory} flow we first mention the assumptions made in deriving Eq.~(\ref{e.Ldot}) and discuss whether these assumptions will hold in our case as well. Firstly, it is assumed that the normal fluid is at rest with respect to the reference frame, {\it i.e.}, in our case with respect to the sphere. This is qualitatively fulfilled because the normal fluid is roughly at rest with respect to the sphere within the viscous penetration depth that is of the same order as the radius at about 1 K while the superfluid moves with respect to the sphere. Secondly, the superflow is assumed to be initially homogeneous. The superfluid velocity field around the sphere varies over a distance of the order of the radius $R$ while the vortex spacing is given by $L^{-1/2}\ll R$, when values of $L_{sat}$ at typical $v_c$ are used \cite{L}. Therefore, this condition is satisfied as well. Finally, some remarks concerning the initial value of $L_0$ are in order. Suppose the drive is increased from zero without any remanent vorticity in the cell ($L_0 = 0$), then $L$ will remain zero, as indicated very impressively by the recent Osaka measurements \cite{Goto}. Typically, in the experiments it is difficult to prepare a helium sample that is completely free of remanent vortices, because even natural radioactivity and cosmic rays are known to produce them \cite{intermitt1}. Therefore, ultimately some strongly trapped vortex loops will be unpinned and $L$ will grow. In addition, the oscillating body may shed new vortices at sharp protuberances where the local flow velocity may be greatly enhanced. All that may occur at velocities larger than $v_c$ and that is why during an initial up sweep of the driving force, $v_c$ can be largely exceeded as shown in Fig.~\ref{f.veloVSforce}. On the other hand, when reducing the drive in the turbulent regime, $L$ is large and $L_0$ is irrelevant. Below 0.5 K, where the switching between turbulent phases and potential flow occurs when approaching $v_c$, we assume that enough vorticity will have survived during the laminar phases, either on the surface of the body or somewhere else in the cell, so that the potential flow will break down again statistically and a new turbulent phase is born. This will continue until too little power from the drive is available to sustain a turbulent phase at all and potential flow will be stable.

The central result of our work is the frequency dependence of $v_c$. This we analyze now on the basis of Eq.~(\ref{e.Ldot}) and Eq.~(\ref{e.Lgrow}) as follows. The time scale $\tau$ determines how fast $L$ can reach the avalanche where the very rapid growth sets in. If $L$ is to reach the avalanche, this must happen on a time scale shorter than one period. Qualitatively, we take the time in which the velocity changes from zero to the maximum, {\it i.e.}, about one quarter of a period. Hence, we have the condition
\begin{equation}
\omega \tau < 1/4,
\label{e.omegatau} 
\end{equation}
and therefore
\begin{equation}
v_s > v_c \approx \sqrt{8\kappa \omega/\beta} = 2.83 \sqrt{\kappa \omega/\beta}.
\label{e.vc}
\end{equation}

\begin{figure}[!th]
\centerline{\includegraphics[width=0.9\columnwidth,clip=true]{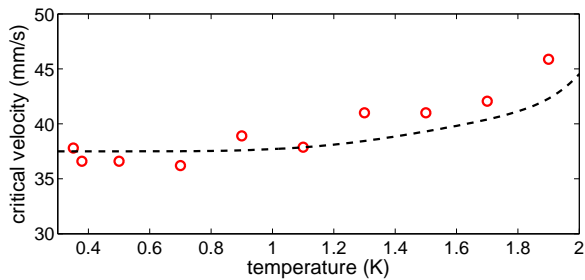}}
\caption{\label{f.vctemp} (color online) Temperature dependence of the critical velocity of a 100 $\mu$m sphere oscillating at 236 Hz. The variation due to $\beta(T)$ is indicated by the dashed line.} 
\end{figure}

To compare Eq.(\ref{e.vc}) with the experimental data we evaluate $\beta$ for $^4$He by referring to tabulated values of $\alpha$ and $\alpha^{\prime}$ \cite{Russ}. Setting the constants $A,B$ = 1, we find $\beta$ = 0.95 (at 1.3 K), 0.89 (at 1.6 K), and 0.79 (at 1.9 K). Below 1 K mutual friction is very small, hence we set $\beta$=1. This implies a slow increase of $v_c$ by about 10\% in qualitative agreement with experimental results as displayed in Fig.~\ref{f.vctemp}.

Note that the source driving the oscillatory flow, in particular the geometry, the size, and the surface properties of the oscillating body, do not appear in Eq.~(\ref{e.vc}), {\it i.e.}, we have a true universal intrinsic critical velocity below which turbulence cannot exist in oscillatory flow.

We have also investigated numerically the evolution of the vortex density in oscillatory flow. A simple first choice might be to let $v_s$ in Eq.~(\ref{e.Ldot}) vary harmonically. This, however, periodically changes the sign of the growth term which obviously is incorrect. Alternatively, one could try $v_s(t) \sim |\sin(\omega t)|$, where the modulus is used to keep the correct sign of the growth term. But this is a unidirectional motion and not an oscillation. It is only useful to confirm that the initial value $L_0>$ 0 is irrelevant once $L$ has begun to grow and that $L(t)$ can follow the time dependent velocity field $v_s(t)$ only if condition (\ref{e.omegatau}) is fulfilled, otherwise the oscillations of $L$ are much reduced and $L_{sat}$ cannot be reached any more. Obviously, this approach is too simple and a rigorous extension of Eq.~(\ref{e.Ldot}) to the oscillating case is needed but, to our knowledge, is presently not available.

\section{Simulations}\label{simulations}
We have also simulated the oscillations of a sphere in the superfluid at zero temperature using the vortex filament model. Assuming a smooth sphere, the calculations with a R=100 $\mu$m sphere indicate a critical velocity of order 100 mm/s at 200 Hz (see also Ref. \cite{Vinen}), which is about 2$\ldots$3 times higher than the experimental one. Those time demanding simulations are still far from being complete. Preliminary calculations indicate that the critical velocity goes up with frequency. Also the calculations for temperature 1.5~K, where the normal flow is taken as Stokes flow, predict that the temperature dependence of the critical velocity is weak, like estimated from the $\beta$ term. It is clear that a single vortex ring in homogeneous oscillating flow is stable. Therefore the boundaries and inhomogeneous flow around the object are the main reasons why the initial (remanent) vortices go unstable. But this does not mean that there cannot be a ``vortex avalanche" later when the line density has grown large enough containing enough vortices of different size, which is the situation on which Kopnin's model is based. It is also true that the simulations indicate that the initial increase in vortex density occurs during several periods, but same occurs also with Eq.~(\ref{e.Ldot}) if initially $L_0\lll L_{sat}$. Initially the increase in vortex length is due to vortex stretching and subsequent collision to the sphere, like noted in Ref.~\cite{Goto}. Unfortunately, we have not managed to reach a steady state due to limited computer resources available. The range of scales that can be simulated is also very limited. It is still an open question what is the turbulent state really like around the vibrating objects, whether it is a state where the vortex density is of order $L_{sat}=(v_s/\kappa)^2$ or much smaller. For that, one should take into account the sphere motion self-consistently, like done by Kivotides {\it et al.} \cite{Kivotides}. In principle near $v_c$ the line density should be low, but so far the simulations only indicate either a fast grow (considered as $v>v_c$) or drop to some small line density (with no clear velocity dependence) if we start from configuration containing few hundred vortices. Hysteresis is also seen in simulations. If initially the vortex number is small, and especially with small vortices, the required velocity, above which the turbulence develops, is larger than if started with the configuration with many vortices.  

\section{Conclusions}\label{conclusions} 
 In summary, our results can be understood, at least in a qualitative manner, both by employing the superfluid Reynolds number and by applying Kopnin's theory to oscillating superflow. A more rigorous theoretical description of the onset of turbulence in oscillatory flow is desirable. Hopefully, also further progress in the numerical simulations will improve our understanding of the experimental results. Finally, we note that Eq.~(\ref{e.vc}) can be applied to turbulence in superfluid $^3$He as well, provided the temperatures is low enough so that $\beta>$ 0 can be achieved \cite{Matti} and the oscillation frequency is low enough so that $v_c$ is smaller than the Landau critical velocity for pair breaking.

\begin{acknowledgments}
We are very grateful to J. J\"ager, H. Kerscher, M. Niemetz, and B. Schuderer for their excellent thesis work in the former Low Temperature Group at Regensburg University, where the sphere data were obtained. J. Hosio kindly informed us about his result with a tuning fork. R.H. acknowledges the support from the Academy of Finland (Grant 114887) and appreciates  discussions with M. Tsubota, H. Yano, M. Kobayashi, and S. Fujiyama. W.S. has benefited very much from his visits to the Helium Group lead by M. Krusius of the Low Temperature Laboratory at Helsinki University of Technology that were made possible by the ULTI program of the EU. We both thank H. Yano very much for showing us his new results on the frequency dependence of the critical velocities of his vibrating wires before publication. Helpful discussions with N.B. Kopnin and G.E. Volovik are most gratefully acknowledged.
\end{acknowledgments}


\begin{thebibliography}{}
\bibitem{L}
Earlier and much smaller estimates of $L$ in Refs.\cite{intermitt1,intermitt2} were based only on dimensional arguments.
\end{thebibliography}


\end{document}